\documentstyle[preprint,pra,aps]{revtex}

\def\ket#1{|#1\rangle} \def\bra#1{\langle#1|}

\tightenlines

\begin{document} 
\draft 
\title{Error correction for mutually interacting qubits} 
\author{Julio Gea-Banacloche} 
\address{Department of
Physics, University of Arkansas, Fayetteville, AR 72701} 
\date{\today}
\maketitle 
\begin{abstract}  For a simple model of mutually interacting qubits
 it is shown how the errors induced by mutual interactions can be eliminated  
 using concatenated coding.  The model is solved exactly for arbitrary interaction 
 strength, for two well-known codes, one and two levels deep: this allows one to see
 under which circumstances error amplitudes add coherently or incoherently.
 For deeper concatenation, approximate
 results are derived which make it possible to calculate
 an approximate ``threshold'' value for the product of 
 interaction strength and free evolution time, below which the failure probability
 for an encoded qubit decreases exponentially with the depth of the encoding.  
 The results suggest that concatenated coding could fully handle the 
 errors arising from mutual interactions, at no extra cost, in terms of 
 resources needed, from what would be required to deal with random environmental 
 errors. 
\end{abstract}

\narrowtext

\section{Introduction and model}
In previous papers \cite{myself1,myself2} I have discussed how unwanted mutual 
interactions between qubits
can be a source of error in a quantum computer (or quantum memory register) and 
studied some of the characteristic features of this type of error.  It was found in
\cite{myself1} that, if
the characteristic interaction energy is $\hbar\delta$, the fidelity ${\cal F}$
of the register 
decreases with time $t$ and with the total number of qubits $N$ as
\begin{equation}
{\cal F} \sim e^{-N(\delta t)^2}
\label{zeroa}
\end{equation}
Equation (\ref{zeroa}) suggests that $(\delta t)^2$ can be interpreted as a sort of 
failure probability per qubit.  
In \cite{myself2} it was shown that the general scaling (\ref{zeroa})
with $t$ holds even when the register is being acted on by ``gates'' such as one might
use in the course of a quantum computation: that is, the error accumulates 
quasi-coherently in time, and there is no ``quantum Zeno effect'' associated with normal 
gate operation.  (This suggests, in particular, that it is not necessary to simulate a 
full computation to estimate the magnitude of this effect; instead, it should be enough
to calculate the survival probability of an appropriate initial state of the register
under the action of the appropriate Hamiltonian \cite{garg}.)

Mutual interactions are Hamiltonian and hence, in principle, reversible, and there exists,
in fact, a variety of special techniques 
for eliminating, or greatly reducing, these errors: in particular,
techniques similar to, and inspired by, the use of refocusing pulses in magnetic 
resonance \cite{lloyd,cory}.  Nonetheless, it is worthwhile 
to look at the question of how well these 
errors could be eliminated by the use of ordinary error-correction methods based on 
quantum error-correcting codes \cite{preskill,knill1}, 
and in particular on concatenated coding, if only because one 
would like to be able to make an informed choice between various available methods.
Moreover, although pulse methods could in principle be used to deal with certain kinds
of environmental noise, it seems very likely that, in large-scale implementations of
quantum computers, error-correcting codes will have to be used, in any event, to deal with at
least some of these random environmental errors.  Assuming, then, that these codes would
already be in place, it is natural to ask how much one could get from them. 

While there is no question that the present theory of quantum error-correction codes 
covers mutually interacting qubits as a special case, this is not a case that is often 
explicitly discussed, and, in particular, it is not immediately clear whether, in order to 
obtain a threshold estimate for these kinds of errors one should add them coherently 
(adding error amplitudes) or incoherently (adding error probabilities).  This is one 
of the issues explored in this paper, for the two best-known quantum error-correcting 
codes: the 5-qubit and 7-qubit codes, and their concatenated forms.  The conclusion, at least 
for the kind of interaction Hamiltonian assumed, is that these codes could fully handle the 
errors arising from mutual interactions, almost certainly at no extra cost in terms of 
resources needed (e.g., depth of concatenation). This is an encouraging result which may not
have been immediately obvious before.

The interaction to be considered here is one that results in pure phase shifts, conditioned on 
the relative state of neighboring qubits, according to a Hamiltonian
\begin{equation}
H = \sum_{n=1}^{N-1} \hbar\delta \sigma_{zn}\sigma_{z,n+1} 
\label{one}
\end{equation}
Examples of physical systems leading to interactions of this form have been discussed 
in previous publications \cite{myself1,myself2}; in \cite{myself2} it has also been shown that, 
for instance, changing 
$\sigma_z$ to $\sigma_x$ (bit errors instead or phase errors) does not modify substantially
the scaling (\ref{zeroa}).

In order to simplify the calculation, the (unrealistic) assumption is made that the 
qubits are divided into physical blocks, each of length $N_c$, which do not interact with 
each other; that is, the interaction Hamiltonian is applied separately to each such block. 
The blocks are intended to represent the ``logical'' qubits in a computer encoded for 
error correction.  When the code is concatenated, the decoupling assumption is applied at 
the lowest (deepest) level:  For instance, for a code concatenated once, a logical qubit 
is made of $N_c\times N_c$ physical qubits, and it is assumed that the $N_c$ lowest-level 
blocks do not interact with each other.  While unrealistic, it is expected that this 
approximation should not modify the results very much, based on the numerical calculations 
made using a similar ``decoupling of registers'' approximation in ref. \cite{myself2}.

In the following Section, the action of this Hamiltonian on a register encoded for error 
correction is analyzed, and then in Section III the analysis is extended to a concatenated 
 code.  The goal is to derive an expression for the ``failure probability'' 
per qubit, that is to say, the probability that the evolution under the Hamiltonian 
(\ref{one}) may lead to a state which cannot be recovered by the error-correction 
protocol.  Section II shows a few general features of this failure probability, such as
its dependence on the initial state considered, and the way some error amplitudes add
coherently and some do not, depending on the code considered.  These results pave the way for
the more complicated situation discussed in Section III.  It is to be noted that the failure
probability is calculated here for arbitrarily large interaction strength, for up to
two-level-deep concatenated codes.  

\section{Results for simple distance 3 codes}

Distance 3 codes can correct any of the three basic one-qubit errors represented by the 
three Pauli matrices $\sigma_x$, $\sigma_y$, and $\sigma_z$.  They will typically fail 
whenever two errors happen simultaneously in the same logical block.

The Hamiltonian (\ref{one}) causes two phase errors in each physical block every time 
it is applied, but the actual damage it does to an encoded state turns out to depend on the 
code used, and on the state it acts upon.  Two examples are considered in this section.

The evolution operator resulting from (\ref{one}) is, for a single logical block,
\begin{equation}
U(t) = \prod_{n=1}^{N_c-1} \left(\cos\delta t + i \sigma_{zn}\sigma_{z,n+1} 
\sin\delta t\right) 
\label{two}
\end{equation}
since the Pauli matrices for different qubits commute.  Thus, a simple perturbative 
estimate of the failure probability would be as follows:  if it is assumed that each term 
of the form $\sigma_{zn}\sigma_{z,n+1}$ leads to a distinct (orthogonal) uncorrectible 
error, there are $N_c-1$ such terms in (\ref{two}), and the sum of the squares of their 
amplitudes is
\begin{eqnarray}
P_{pert}(t) &&= (N_c-1)\left(\cos^2 \delta t\right)^{N_c-1} \sin^2 \delta t
\nonumber\\ 
&&\simeq (N_c-1) (\delta t)^2 \label{three}
\end{eqnarray}
for sufficiently small interaction strength and time $\delta t$.  

As it turns out, this expression is not, in general, correct, even for small $\delta t$,
because of two assumptions:  that all terms of the form $\sigma_{zn}\sigma_{z,n+1}$ lead 
invariably to fatal (uncorrectible) errors, and that the different errors are orthogonal.
This is explored in detail in what follows.

\subsection{The 7-qubit code}

Consider the 7-qubit code introduced by Steane \cite{steane}:
\begin{mathletters} 
\label{four} 
\begin{eqnarray}
\ket{0_L} = {1\over\sqrt 8} \bigl(&&\ket{0000000}+ \ket{0001111} +
\ket{0110011}  \nonumber \\ &&+ \ket{0111100}+ \ket{1010101} +
\ket{1011010} \nonumber \\ &&+\ket{1100110} + \ket{1101001} \bigr)
\label{foura} 
\end{eqnarray} 
\begin{eqnarray} \ket{1_L} = {1\over\sqrt 8}
\bigl(&&\ket{1111111}+ \ket{1110000} + \ket{1001100} 
\nonumber \\
&&+\ket{1000011} + \ket{0101010} + \ket{0100101} \nonumber \\ &&+
\ket{0011001} + \ket{0010110} \bigr) 
\label{fourb} 
\end{eqnarray}
\end{mathletters} 
Because the code is non-degenerate, all 7 states of the form $\sigma_{zn}\ket{0_L}$ 
(with $n=1,\dots,7$) are orthogonal to each other and to $\ket{0_L}$.  Altogether, then,
these eight states form an orthogonal basis of the space spanned by the eight 
kets appearing in (\ref{foura}).  But then, since any combination of products of 
$\sigma_{zn}$ acting on $\ket{0_L}$ does nothing but to change the relative signs in the 
superposition of these eight kets, it follows that it can always be written as a 
superposition of ``one-error'' states:
\begin{equation}
U(t)\ket{0_L} = C_0\ket{0_L} + \sum_{n=1}^7 \sigma_{zn} \ket{0_L}
\label{five}
\end{equation}
and this is always correctible.  A similar reasoning applies to the state $\ket{1_L}$, 
which means that, if the initial state of the system is either $\ket{0_L}$ or $\ket{1_L}$, 
the evolution operator (\ref{two}) does not introduce any fatal errors at all.

On the other hand, the initial state of the system is much more likely to be a coherent 
superposition of $\ket{0_L}$ and $\ket{1_L}$, and for such states the situation is 
different, because, although error correction will restore $\ket{0_L}$ and $\ket{1_L}$ 
separately, it will also introduce a phase difference between them.  The following 
useful relations can be easily established from the explicit forms (\ref{four}):  
\begin{mathletters} 
\label{six} 
\begin{eqnarray}
\sigma_{z1}\sigma_{z2}\ket{0_L} = \sigma_{z5}\sigma_{z6}\ket{0_L} &&= -\sigma_{z3}\ket{0_L} 
\nonumber \\
\sigma_{z2}\sigma_{z3}\ket{0_L} = \sigma_{z4}\sigma_{z5}\ket{0_L} = 
\sigma_{z6}\sigma_{z7}\ket{0_L} &&= -\sigma_{z1}\ket{0_L} \nonumber \\
\sigma_{z3}\sigma_{z4}\ket{0_L} &&= -\sigma_{z7}\ket{0_L}
\label{sixa} 
\end{eqnarray} 
\begin{eqnarray}
\sigma_{z1}\sigma_{z2}\ket{1_L} = \sigma_{z5}\sigma_{z6}\ket{1_L} &&= \sigma_{z3}\ket{1_L} 
\nonumber \\
\sigma_{z2}\sigma_{z3}\ket{1_L} = \sigma_{z4}\sigma_{z5}\ket{1_L} = 
\sigma_{z6}\sigma_{z7}\ket{1_L} &&= \sigma_{z1}\ket{1_L} \nonumber \\
\sigma_{z3}\sigma_{z4}\ket{1_L} &&= \sigma_{z7}\ket{1_L}
\label{sixb} 
\end{eqnarray} 
\end{mathletters} 
and it is easily seen from this that when error correction is applied to the state 
$\sigma_{zn}\sigma_{z,n+1} (\alpha\ket{0_L}+\beta\ket{1_L})$, the result will be the state 
$(-\alpha\ket{0_L}+\beta\ket{1_L})$, that is, a state with the wrong relative phase.  
Equations (\ref{six}) also show that the assumption that different 
$\sigma_{zn}\sigma_{z,n+1}$ lead to orthogonal error states is incorrect:  for instance,
$\sigma_{z1}\sigma_{z2}$ and $\sigma_{z5}\sigma_{z6}$ lead to the same error state.

Using Eqs. (\ref{six}) it is straightforward to derive the correct form of the 
approximate failure probability for small $\delta t$.  When only the terms of first 
order in $\delta t$ are kept in the expression (\ref{two}), acting on a general state 
$\alpha\ket{0_L}+\beta\ket{1_L}$, one obtains 
\begin{equation}
\delta t\,(3\sigma_{z1}+2\sigma_{z3}+\sigma_{z7})(-\alpha\ket{0_L}+\beta\ket{1_L})
\label{seven}
\end{equation}
which means that the probability that after error correction one might end up with the 
state $-\alpha\ket{0_L}+\beta\ket{1_L}$ is $(3^2+2^2+1^2)(\delta t)^2 = 14 (\delta t)^2$. 
One might directly call this the failure probability (since the state 
$-\alpha\ket{0_L}+\beta\ket{1_L}$ is generally not equal to the initial state 
$\alpha\ket{0_L}+\beta\ket{1_L}$) or define the failure probability  as including 
explicitly the square of the overlap between the initial and the final state 
(see Eq. (\ref{deffailprob}), below, for the precise definition), in which 
case one finds
\begin{equation}
P \simeq 14 (\delta t)^2 \left(1-\left(|\alpha|^2-|\beta|^2\right)^2 \right)
\label{eight}
\end{equation}
where the $\simeq$ sign is a reminder that this is only valid for small $\delta t$.  
The failure probability (\ref{eight}) is largest for the symmetric linear superpositions 
$(\ket{0_L}\pm\ket{1_L})/\sqrt{2}$.  Note how it is different from the 
result (\ref{three}) (for $N_c=7$), obtained under the assumption that the errors 
generated by different $\sigma_{zn}\sigma_{z,n+1}$ were all orthogonal, and it is also 
not equal to the same expression with $N_c-1$ replaced by $(N_c-1)^2$, which 
would correspond to a ``worst case scenario'' in which all the different errors 
added coherently.  The coefficient $14$ falls instead in between the two extremes of 
6 $(N_c-1)$ and 36 ($(N_c-1)^2$).

The definition of the failure probability $P$ used above is 
\begin{equation}
P=1-\left|\bra{\Psi_0}U(t)\ket{\Psi_0}\right|^2 - \sum_{n=1}^{N_c} 
\left|\bra{\Psi_0}\sigma_{z,n}U(t)\ket{\Psi_0}\right|^2
\label{deffailprob}
\end{equation}
that is, one minus the total probability that, after a time $t$, the system may be 
found in 
either the original state $\ket{\Psi_0}$, or one of the $N_c$ correctible states 
$\sigma_{z,n}\ket{\Psi_0}$ (there are, of course, other correctible states, of the 
form $\sigma_{x,n}\ket{\Psi_0}$ and $\sigma_{y,n}\ket{\Psi_0}$, but they cannot be 
accessed by this interaction operator).

Using this definition it is relatively straightforward to calculate $P$ exactly for 
arbitrarily large $\delta t$.  The following relation is useful:
\begin{mathletters}
\label{ten}
\begin{equation}
\bra{0_L} \sigma_{z,n_1},\ldots, \sigma_{z,n_k} U \ket{0_L} = 
(-1)^k\bra{1_L} \sigma_{z,n_1},\ldots, \sigma_{z,n_k} U \ket{1_L}
\label{tena}
\end{equation}
\begin{equation}
\bra{0_L} \sigma_{z,n_1},\ldots, \sigma_{z,n_k} U \ket{1_L} = 0
\label{tenb}
\end{equation}
\end{mathletters}
where $k=0,\ldots, 7$. This follows directly 
from the fact that $U$ involves always the product of an even number of $\sigma_z$ 
operators, and that the code words used in $\ket{0_L}$ are the complementaries of
the ones used in $\ket{1_L}$.

For future use it is convenient to define
\begin{mathletters}
\label{ten1}
\begin{equation}
f_0  = \left| \bra{0_L}  U \ket{0_L} \right|^2 = \left| \bra{1_L}  U \ket{1_L} \right|^2 
\label{ten1a}
\end{equation}
\begin{equation}
f_1  = \sum_{n=1}^7\left| \bra{0_L} \sigma_{z,n} U \ket{0_L} \right|^2 = \sum_{n=1}^7
\left| \bra{1_L}  U \ket{1_L} \right|^2
\label{ten1b}
\end{equation}
\end{mathletters}

Direct calculation then yields
\begin{equation}
f_0 = 1-f_1 = \cos^2(\delta t) \cos^2(2\delta t) \cos^2(3 \delta t)
\label{ten2}
\end{equation}
and
\begin{equation}
P(t) = f_1  
\left(1-\left(|\alpha|^2-|\beta|^2\right)^2 \right)
\label{eleven}
\end{equation}
and it is easy to verify that (\ref{eight}) is the lowest-order term in the 
expansion of (\ref{eleven}) in powers of $\delta t$.

The failure probability (\ref{eleven}) for the case $|\alpha|=|\beta|$ is plotted in 
Figure 1 (solid line) as a function of $\delta t$, from $\delta t = 0$ to $\pi/2$. 
From $0$ to $\pi$, the curve is symmetric about $\pi/2$, as is to be expected from the
form of the evolution operator (\ref{two}).  

\subsection{The 5-qubit code}

In 1996 Laflamme et al. \cite{laflamme}
 introduced a 5-qubit code which can correct general one-qubit 
errors.  A possible encoding is   
\begin{mathletters} 
\label{twelve} 
\begin{eqnarray}
\ket{0_L} = {1\over\sqrt 8} \bigl(&&-\ket{00000}+ \ket{01111} -
\ket{10011} + \ket{11100}  \nonumber \\ &&+ \ket{10101} +
\ket{11010} +\ket{00110} + \ket{01001} \bigr)
\label{twelvea} 
\end{eqnarray} 
\begin{eqnarray} \ket{1_L} = {1\over\sqrt 8}
\bigl(&&-\ket{11111}+ \ket{10000} + \ket{01100} -\ket{00011} 
\nonumber \\
&&- \ket{01010} - \ket{00101} + \ket{11001} + \ket{10110} \bigr) 
\label{twelveb} 
\end{eqnarray}
\end{mathletters} 
The codewords can be obtained from the 7-qubit code by deleting any two qubits 
(in this case, the first two), but the pattern of signs is nontrivial.

Note that the logical 0 and 1 states are still superpositions of eight basis vectors
each (where the basis vectors are simultaneous eigenstates of all the $\sigma_{z,n}$);
 hence, the 
ten one-error states $\sigma_{z,n}\ket{0_L}$ and $\sigma_{z,n}\ket{1_L}$, plus the two 
no-error states $\ket{0_L}$ and $\ket{1_L}$ are not enough to span the space resulting 
from all the possible combinations of the 16 basis kets.  As a result, there are now 
no states which exhibit fully correctible evolution (unlike the $\ket{0_L}$ and 
$\ket{1_L}$ did for the 7-qubit code).  However, we still have the 
phenomenon of non-orthogonal errors.  In particular, we find that
\begin{mathletters} 
\label{thirteen} 
\begin{equation}
\sigma_{z2}\sigma_{z3}\ket{0_L} = 
\sigma_{z4}\sigma_{z5}\ket{0_L}  
\label{thirteena} 
\end{equation} 
\begin{equation}
\sigma_{z2}\sigma_{z3}\ket{1_L} = 
\sigma_{z4}\sigma_{z5}\ket{1_L}  
\label{thirteenb} 
\end{equation} 
\end{mathletters} 
and neither of these expressions is equivalent to a linear combination of single-qubit 
errors acting on $\ket{0_L}$ or $\ket{1_L}$; in fact, they are orthogonal to all such 
states, and to the no-error states $\ket{0_L}$ and $\ket{1_L}$ as well.  

The calculation of $P(t)$, as given by the definition (\ref{deffailprob}) is a little 
more involved, although the results (\ref{ten}) still hold.  
Defining as before
\begin{mathletters}
\label{thirteen1}
\begin{equation}
f_0  = \left| \bra{0_L}  U \ket{0_L} \right|^2 = \left| \bra{1_L}  U \ket{1_L} \right|^2 
\label{thirteen1a}
\end{equation}
\begin{equation}
f_1  = \sum_{n=1}^5\left| \bra{0_L} \sigma_{z,n} U \ket{0_L} \right|^2 = \sum_{n=1}^5
\left| \bra{1_L}  U \ket{1_L} \right|^2
\label{thirteen1b}
\end{equation}
\end{mathletters}
one finds, thanks to Eqs.~(\ref{ten}), that $P(t)$ can still be written as
\begin{equation}
P(t) = 1-f_0 - f_1 \left(|\alpha|^2-|\beta|^2\right)^2
\label{thirteen2}
\end{equation}
but this time, unlike for the 7-qubit code, one does not have $f_0=1-f_1$.  Instead,
\begin{mathletters}
\label{thirteen3}
\begin{equation}
f_0 = \cos^4(\delta t) \cos^2(2\delta t)
\label{thirteen3a}
\end{equation}
\begin{equation}
f_1 = \sin^2(\delta t) \left(1+\cos^2(\delta t) 
\cos^2(2 \delta t) \right)
\label{thirteen3b}
\end{equation}
\end{mathletters}
The failure probability (\ref{thirteen2}) for the case $|\alpha|=|\beta|$ is plotted in 
Figure 2 (solid line) as a function of $\delta t$, from $\delta t = 0$ to $\pi/2$. 
For small values of $\delta t$ Eq.~(\ref{thirteen2}) reduces to
\begin{equation}
P \simeq (\delta t)^2 \left[6-2 \left(|\alpha|^2-|\beta|^2\right)^2 \right]
\label{fifteen}
\end{equation}

The reason $f_0\ne 1-f_1$ for the 5-qubit code is that, as mentioned above, 
 the states $\ket{0_L}$ and 
$\sigma_{z,n}\ket{0_L}$ do not form a complete basis of the space spanned by the 
eight kets which make up $\ket{0_L}$.  For the 7-qubit code, instead, they do, and, since 
the operator $U$ acts entirely within that space 
(when the initial state is $\ket{0_L}$), 
any part of $U\ket{0_L}$ which is not accounted for by the original state $\ket{0_L}$
has to show up in one of the $\sigma_{z,n}\ket{0_L}$ (and the same for the initial state 
$\ket{1_L}$).   

\section{Concatenated codes}

A relatively straightforward way to obtain a code that can correct more than one-qubit 
errors is to concatenate a one-error correcting code, in principle, as many times (or 
levels deep) as necessary \cite{preskill,knill1}.  
In a code concatenated once, each logical qubit consits of 
$N_c$ blocks of $N_c$ (lower-level) qubits each.  This code can protect 
against general errors in any three lower-level qubits, and also against many 
other combinations of multiple lower-level errors.  The minimum number of lower-level 
errors required to cause the code to fail is four, two in one upper level block
(causing that whole block to fail) and two in a different one.

In this section I look first at the 7-qubit and 5-qubit codes concatenated once.  A physical 
(lowest-level) qubit will be identified by two indices, $n$ and $m$, running from $1$ to
$N_c$:  the first index denotes the upper-level block it belongs to, 
and the second index the qubit's 
position within that block.  With the assumption that different blocks do not 
interact, the time-evolution operator (\ref{two}) becomes
\begin{equation}
U(t) = \prod_{n=1}^{N_c}\prod_{m=1}^{N_c-1} \left(\cos\delta t + i \sigma_{nm}\sigma_{n,m+1} 
\sin\delta t\right) 
\label{seventeen}
\end{equation}
From now on, for simplicity, the subscript $z$ on $\sigma_z$ will be dropped, as only 
$\sigma_z$ is used.

\subsection{The 7-qubit code}

If all the errors generated by the individual terms in (\ref{seventeen}) were orthogonal, 
it would be a relatively simple matter to calculate the probability of an uncorrectible 
error to lowest order.  It would be given by the terms of the form $\sigma_{n_1m_1}
\sigma_{n_1,m_1+1}\sigma_{n_2m_2}\sigma_{n_2,m_2+1}$, with $n_1\ne n_2$ and $m_1, m_2$ 
arbitrary.  There are $C(N_c,2)$ ways to choose the indices $n_1$ and $n_2$, and for 
each choice there are $(N_c-1)^2$ different ways to choose $m_1, m_2$, so there is a 
total of
\begin{equation}
{N_c\choose2}
\left(N_c-1 \right)^2 
\label{eighteen}
\end{equation}
terms in (\ref{seventeen}) which contribute to the failure probability in lowest order,
and all these terms are proportional to $(\delta t)^4$ (after squaring the probability 
amplitudes).

However, as has been shown already in the previous section, not all the terms enumerated 
in (\ref{eighteen}) are mutually orthogonal.  For the 7-qubit code, according to 
(\ref{six}), there are two sets of values for $m_1$ (one set of three values, 
and one set of two values) which result in identical states.  
When the equivalent choices for $m_2$ are factored in and added, and 
the probability amplitudes of the corresponding states are squared, one finds, instead 
of the factor $(N_c-1)^2 = 36$ in (\ref{eighteen}), a factor
\begin{equation}
(3^2+2^2+1^2)(3^2+2^2+1^2) = 14^2 = 196
\label{nineteen}
\end{equation}
As a result, to lowest order, the failure probability for this two-level code is given by 
\begin{eqnarray}
P^{(2)} &&\simeq 196
{7\choose2}
(\delta t)^4 \left(1-\left(|\alpha|^2-|\beta|^2\right)^2 \right)
\nonumber \\
&&= 4116 \, (\delta t)^4 \left(1-\left(|\alpha|^2-|\beta|^2\right)^2 \right)
\label{twenty}
\end{eqnarray}
for a state $\alpha\ket{0_L}+\beta\ket{1_L}$. The dependence on $\alpha$ and 
$\beta$ exhibited by (\ref{twenty}) may not be not immediately obvious at this point 
(it will be derived later), but it shows that, 
for the concatenated code also, the individual states $\ket{0_L}$ and $\ket{1_L}$ are 
fully correctible.  

Note that the maximum error probability $P_{max}^{(2)}$ for the once-concatenated code,
according to (\ref{twenty}), is related to the corresponding one for the plain code, 
(\ref{eight}), by the formula 
\begin{equation}
P_{max}^{(2)} =  {7\choose2} \left(P_{max}^{(1)}\right)^2
\label{twenty1}
\end{equation}
and the right-hand side of (\ref{twenty1}) is simply the probability that two blocks
might fail {\it independently}, times the number of ways to choose two blocks.  This is
because, to this order (recall that these are only approximate results for small $\delta t$),
errors across different blocks are orthogonal, ie., they add incoherently.  This can also be
seen from Equation (\ref{nineteen}):  within each block there is some coherent addition of
error amplitudes (which results in the terms $3^2$ and $2^2$), but then the total error 
probabilities for each block are simply multiplied.

This is an important result which generalizes to higher orders, as discussed later in this
Section (subsection C). The key point is that all the errors which would cause a 
particular block to fail
are orthogonal to all the errors that would cause a {\it different} block
to fail, {\it to lowest order}.  This is because to lowest order all it takes is two errors
in one block to cause that block to fail, and because the original code is nondegenerate,
which means that all expectation values of the form $\bra{\xi_L}\sigma_i\sigma_j\ket{\eta_L}$
(where $\eta, \xi = 0,1$) within the same block are zero.  

Equation (\ref{twenty1}) does not, however, hold for arbitrarily large $\delta t$, when
multiple (more than two) errors within the same block need to be taken into account.  
For a code concatenated only once it is still possible, albeit very cumbersome, to derive this 
failure probability for arbitrary $\delta t$.  The definition (\ref{deffailprob}) 
has to be changed to
\begin{equation}
P=1-\left|\bra{\Psi_0}U(t)\ket{\Psi_0}\right|^2 - \sum_n 
\left|\bra{\Psi_0}A_nU(t)\ket{\Psi_0}\right|^2
\label{twentyone}
\end{equation}
where the $\{A_n\}$ represent a maximal set of correctible error operators leading to 
orthogonal states. The difficulty is that now, unlike in the previous Section, not all 
possible correctible errors lead to orthogonal states.  For instance, any number of 
errors within one single block are now correctible, but certainly not all of these
lead to orthogonal states.  Thus, even though we started out with a 
nondegenerate code, {\it the concatenated 
code is degenerate\/} \cite{gottesman}.  

As far as I can tell, the set $\{A_n\}$ in (\ref{twentyone}) must be constructed by inspection. 
For the 
7-qubit code, concatenated once, I find that the maximum number of 
correctible errors acting on a single block and leading to orthogonal states is 15, 
which can be chosen to be 
\begin{eqnarray}
\sigma_{nm}\qquad (m=1,\ldots,7) \nonumber \\
\sigma_{n1}\sigma_{nm} \qquad (m=2,\ldots,7) \nonumber \\
\sigma_{n2}\sigma_{n3} \nonumber \\
\sigma_{n1}\sigma_{n4}\sigma_{n5}
\label{twentytwo}
\end{eqnarray}  
Of these, the first 7 are, of course, fully correctible at the lower level, 
whereas the others, upon correction at the lower level, lead to what look like 
``block'' phase errors at the higher level.

The single-block errors (\ref{twentytwo}) can be combined in a large number of ways 
among the different blocks to 
yield other, orthogonal, correctible errors.  Fortunately, however, the actual number of 
such combinations which make a nonvanishing contribution to (\ref{twentyone}) 
is relatively small.
To see this, divide the error
operators acting on a single block into those involving an even number of $\sigma$'s 
(denoted as $E_n$ if they act on the $n$-th block) and those involving an odd 
number of $\sigma$'s (denoted as $O_n$).  Then, for a general multiblock error of 
the form $A=E_{n_1}\ldots E_{n_k}O_{m_1}\ldots O_{m_k}$, 
 Eqs.~(\ref{ten}) for a single block imply that the
expectation value 
\begin{equation} 
\bra{0_L}O_{n_1}\ldots O_{n_k}E_{m_1}\ldots E_{m_l}U\ket{0_L}
\label{twentythree}
\end{equation}
will only be nonzero if the corresponding combination 
$\bra{0_L}\sigma_{n_1}\ldots\sigma_{n_k}\ket{0_L}$ is nonzero for the single, 
non-concatenated, code.  The same is true for the state $\ket{1_L}$, and there are
no cross-terms between $\ket{0_L}$ and $\ket{1_L}$ (also as a direct consequence of
the single-block Eqs.~(\ref{ten})).  Thus, to evaluate (\ref{twentyone}),
it is sufficient to identify the combinations of numbers $\{n_1,\ldots,n_k\}$ which lead 
to a nonzero  $\bra{0_L}\sigma_{n_1}\ldots\sigma_{n_k}\ket{0_L}$ for the 7-qubit 
code, and put odd errors, chosen from the list (\ref{twentytwo}), in the 
 blocks $n_1,\ldots,n_k$, augmented perhaps with even errors, 
always chosen from the list 
(\ref{twentytwo}), in any other blocks, provided, of course, 
that the total error be correctible.  This quickly reduces the total of different 
possibilities to a manageable number:  for instance, since an even error in one block
would cause that block to fail, one cannot have even errors in more than one block at a 
time.  

For reference, for the 7-qubit code, the sets of values of $n_1,\ldots,n_k$ to be used 
are:  $(1,2,3)$, $(1,4,5)$, $(1,6,7)$, $(2,4,6)$, $(2,5,7)$, $(3,4,7)$, $(3,5,6)$, 
the complementary set ($(4,5,6,7)$, $(2,3,6,7)$, and so on), and the set $(1,2,3,4,5,6,7)$.
Then, possible error operators $A_n$ 
to be used in (\ref{twentyone}) might be products such as 
such as $\sigma_{1,m_1}\sigma_{2,m_2}\sigma_{3,m_3}$, or 
$\sigma_{1,m_1}\sigma_{2,m_2}\sigma_{3,m_3}\sigma_{4,1}\sigma_{4,m_4}$,
or $\sigma_{n,1}\sigma_{n,m}$.  In terms of the functions $f_0$, $f_1$ defined in 
(\ref{ten1}), the term(s) $\sum_{m_1,m_2,m_3} |\bra{0_L}
\sigma_{1,m_1}\sigma_{2,m_2}\sigma_{3,m_3}\ket{0_L}|^2$ contribute a term $f_0^4 f_1^3$ to
the expression (\ref{twentyone}); the term $\sum_{m_1,m_2,m_3,m_4} |\bra{0_L} 
\sigma_{1,m_1}\sigma_{2,m_2}\sigma_{3,m_3}\sigma_{4,1}\sigma_{4,m_4}\ket{0_L}|^2$ 
contributes a term $f_0^3 f_1^4$ (since the product of two $\sigma$'s in the same block is
equivalent, except for a sign, to a single $\sigma$ acting on the appropriate state, cf. 
Eqs.~(\ref{six})), and so on.

Collecting all the terms, and keeping track carefully of which ones have opposite
signs in the states $\ket{0_L}$ and the states $\ket{1_L}$, one finds
\begin{equation}
P^{(2)}(t) = \left(1-f_0^7-7f_0^6 f_1 - 28 f_0^4 f_1^3 - 7 f_0^3 f_1^4 - 21 f_0^2 f_1^5
\right) 
\left(1-\left(|\alpha|^2-|\beta|^2\right)^2 \right)
\label{twentyfour}
\end{equation}
with $f_0$ and $f_1$ given by (\ref{ten2}), 
which does reduce to (\ref{twenty}) for small $\delta t$.

The exact result (\ref{twentyfour}) is plotted as a function of $\delta t$ 
in Figure 1 (dashed line), for $|\alpha|=|\beta|$.  It is, perhaps, remarkable that it 
does not look all that different from the single-encoding result (\ref{eleven}), except 
in the very small $\delta t$ region.  It is also clear that Eq. (\ref{twenty1}) does not
hold for large $\delta t$, which means that for large $\delta t$ we do see the effects of 
errors across different blocks adding coherently. 

\subsection{The 5-qubit code}

For the 5-qubit code, one again has to identify the appropriate maximal set of error 
operators leading to orthogonal errors.  For a single block, one finds the following 
set of 15: 
\begin{eqnarray}
\sigma_{nm}\qquad (m=1,\ldots,5) \nonumber \\
\sigma_{n1}\sigma_{nm} \qquad (m=2,\ldots,5) \nonumber \\
\sigma_{n2}\sigma_{nm} \qquad (m=3,\ldots,5) \nonumber \\
\sigma_{n1}\sigma_{n2}\sigma_{nm} \qquad (m=3,\ldots,5)
\label{twentyfive}
\end{eqnarray} 

As in the previous subsection, it is found that only certain combinations of odd errors
across different blocks lead to nonzero contributions to (\ref{twentyone}).  These turn 
out to be much fewer than before, being limited to $n_1,\ldots,n_k = (1,2,5), (1,3,4)$, 
and $(2,3,4,5)$.  

The result, valid for arbitrarily large $\delta t$, turns out to be now
\begin{eqnarray}
P^{(2)}(t) = 1  &&-(f_0^5+5f_0^4(1-f_0)+f_1^4 + 4f_0f_1^3(1-f_1)) \nonumber \\
&&-(4f_0f_1^3+6f_0^2 f_1^2-8f_0^2 f_1^3)\left(|\alpha|^2-|\beta|^2\right)^2
\label{twentysix}
\end{eqnarray}
with $f_0$ and $f_1$ given by Eqs.~(\ref{thirteen3}).  The factors of $1-f_0$ appearing 
in (\ref{twentysix}) come from the sum of the single-block expectation values squared of 
$\sigma_{1}\sigma_{m}$, $m=2,\ldots,5$ and $\sigma_{2}\sigma_{m}$, $m=3,\ldots,5$; 
these seven error operators, acting on the singly encoded $\ket{0_L}$ or $\ket{1_L}$, 
generate the seven orthogonal states necessary to complete a basis of the space spanned 
by the 8 kets appearing in (\ref{twelve}).  The factors of $1-f_1$ come from the sum 
of the single-block expectation values squared of $\sigma_{1}\sigma_{2}\sigma_{m}$,
$m=3,\ldots,5$: these 3 error operators turn out to be the complement of the 5 ones 
appearing in the definition (\ref{thirteenb}) of $f_1$.

The limit of (\ref{twentysix}) for small $\delta t$ is
\begin{equation}
P^{(2)}(t) \simeq (\delta t)^4 \left[360 - 24 \left(|\alpha|^2-|\beta|^2\right)^2\right]
\label{thirty}
\end{equation}
As usual, the failure probability is largest for the symmetric superposition state 
$(\ket{0_L}\pm\ket{1_L})/\sqrt{2}$.  The corresponding coefficient, $360$, can be 
obtained also from the simple arguments used to derive the result (\ref{twenty}) in the
previous subsection:  it equals $C(5,2)(2^2+1^2+1^2)^2$, where the $2^2$ stands for the 
amplitude square of the sum of the two error terms which lead to identical errors, as
shown in Eq.~(\ref{thirteen}) (and the two $1^2$ stand for the other two orthogonal 
errors).

The exact result (\ref{twentysix}) is plotted in Figure 2 (dashed line) as a function of
$\delta t$, for $|\alpha|=|\beta|$.  In this case, unlike for the 7-qubit code, the 
difference with the single-encoding formula (solid line, equation (\ref{thirteen2}))
is substantial; the twice-encoded qubit has a substantially smaller failure probability 
for most values of $\delta t$.  The failure probability actually goes to zero at 
$\delta = \pi/2$, where Eq.~(\ref{seventeen}) yields $U = i\prod_{n=1}^5 \sigma_{n1} 
\sigma_{n5}$.  Upon inspection, it is found to be 
a peculiarity of the concatenated (two-levels deep) 5-qubit code that 
\begin{equation}
\prod_{n=1}^5 \sigma_{n1} \sigma_{n5} \left(\alpha \ket{0_L} + \beta \ket{1_L}
\right) = \sigma_{11} \sigma_{15} \sigma_{22} \sigma_{32} \sigma_{42} \sigma_{52}
\left(\alpha \ket{0_L} + \beta \ket{1_L}
\right)
\label{thirty2}
\end{equation}
and the state on the right-hand side of (\ref{thirty2}) is fully correctible, since it 
amounts to only one block failing.   

\subsection{Results for arbitrarily deep concatenation}

Clearly, the exact calculation of $P^{(n)}$ for a code concatenated $n$ levels deep, 
with $n > 2$, becomes much too cumbersome to be feasible.  However, the results in the 
previous subsections show that the leading term in $\delta t$ can be calculated using 
relatively simple arguments.

Consider, for instance, a code concatenated three levels deep.  It will fail if at least 
two of the high-level qubits fail, but these high-level qubits are themselves 
two-level deep codes.  One can choose the two highest-level qubits that fail in any of
$C(N_c,2)$ ways, and for each of them the number of terms in the expansion of the 
evolution 
operator which lead to a qubit failure are those computed for the previous level.  So, 
for the 7-qubit code, we can immediately iterate (\ref{twenty}) to get
\begin{eqnarray}
P^{(3)}_{max} &&\simeq {7\choose2} \left(4116 (\delta t)^4 \right)^2 \nonumber \\
&&\simeq {7\choose2} \left(P^{(2)}_{max} \right)^2 
\label{thirtyone}
\end{eqnarray}
where the subscript ``max'' indicates that we are looking at the failure probability for
the state that maximizes it (the linear combination $(\ket{0_L}\pm\ket{1_L})/\sqrt{2}$),
and Eq.~(\ref{eight}) has been used for $P^{(1)}$.

Equation (\ref{thirtyone}) contains the essence of a recursion relation which yields the
failure probability at the $n$-th level of concatenation ($n\ge 2$) as
\begin{equation}
P^{(n)}_{max} \simeq {7\choose2}^{-1} \left({7\choose2} P^{(1)}_{max}\right)^{2^{n-1}}
\label{thirtytwo}
\end{equation}
Figure 3 shows the approximate $P^{(n)}$, calculated from (\ref{thirtytwo}), for 
$n=1,\ldots,4$ (solid lines), and also the exact results (\ref{eleven}) and 
(\ref{twentyfour}), for comparison.  Based on these examples, it would seem that 
Equation (\ref{thirtytwo}) overestimates
the failure probability, which would make it a somewhat conservative estimate  

As encoding depth $n$ increases, the failure probability (\ref{thirtytwo}) will, for a 
given $\delta t$ increase or decrease depending on whether $C(7,2) P^{(1)}_{max}$ is 
greater than or less than one.  This yields the ``threshold'' value for $(\delta t)^2$, 
below which virtually error-free operation can be achieved for sufficiently deep 
encoding.  Using (\ref{eight}), one finds the condition:
\begin{equation}
(\delta t)^2 < {1\over 21\cdot 14} = 3.4\times 10^{-3}
\label{thirtythree}
\end{equation}
(Figure 3 shows this as the point where all the curves cross.)
This threshold (\ref{thirtythree}) is comparable to some estimates for the 
correction of independent, random errors by concatenate codes, if one thinks of 
$(\delta t)^2$ as a sort of failure probability for a single physical
qubit.  As pointed out in the Introduction,
this is a natural interpretation (in spite of the fact that two physical qubits 
at a time are modified by this 
interaction), because of the result (\ref{zeroa}).

A similar calculation yields the threshold result for the 5-qubit code (using 
(\ref{fifteen}) and C(5,2)=10): 
\begin{equation}
(\delta t)^2 < {1\over 10\cdot 6} = 1.7\times 10^{-2}
\label{thirtyfour}
\end{equation}

\section{Discussion}

The results in the previous section
suggest that, in principle, the correction of even rather large
undesired interactions between the qubits does not require special methods or 
resources beyond those needed to control independent, random errors arising from the 
interaction of the qubits with the environment. This observation is based, first, 
 on the fact that 
estimates of the threshold for fault-tolerant computation with independent environmental
errors tend to be rather more restrictive than either (\ref{thirtythree}) or 
(\ref{thirtyfour}) \cite{preskill,knill1}; 
and, second, on the fact that for most proposed quantum
computing systems the unwanted interactions betwen qubits would easily satisfy the 
threshold conditions (\ref{thirtythree}) or 
(\ref{thirtyfour}), provided that the time in between error corrections is not 
excessive.

As an example,  
consider a system of nuclear spins, in a solid, for instance (such as in the proposal
by Kane \cite{kane}), separated by $\sim 150$ \AA, and interacting via the magnetic dipole-dipole 
interaction, which would naturally lead to a Hamiltonian like (\ref{one}). For such a
system $\delta$ may be estimated as 
\begin{equation}
\delta \sim {\mu_0 \mu_N^2 \over 4 \pi \hbar d^3} \sim 10^{-2} \hbox{s}^{-1}
\label{thirtyfive}
\end{equation}
Here $\mu_N$ is the nuclear magneton.

Without error correction $\delta t$ becomes already of the order of 1 after 100 s, and
long calculations are impossible.  But assume that one uses error correction and that 
complete error correction of an $n$-order
encoded qubit requires a time $t = c_n \tau_{gate} $, where the time needed to perform 
an elementary logic gate for this system is 
about $\tau_{gate} = 10^{-5}$ s.  Then the threshold is easily reached unless $c_n$ is
exceptionally large, which would make error correction generally impossible anyway
(a large $c_n$ would mean essentially that the computer architecture does not allow
for a large degree of parallelism when correcting errors across different qubits).

Of course, the results presented here have been obtained only for a specific kind of
interaction Hamiltonian and a specific geometry, and would need to be modified for
other cases (e.g., two-dimensional geometries, with more nearest-neighbors per qubit,
or to include non-nearest neighbor interactions).  There is also the simplifying assumption 
made from the outset, of neglecting interactions beween qubits in different blocks.
These complications might require recalculations for specific physical systems and
hardware configurations, but I do not expect them to change substantially the threshold
estimates of the previous Section.  

For instance, for the kind of ``linear'' geometry considered here, including 
interactions between the last qubit of one block and the first one of the next block
would not modify the failure probability to lowest order, since a single two-qubit
error which straddles two blocks does not cause either block to fail; hence, to make any
one block fail, more of these 
errors are required than of the kind of ``internal'' errors
considered in this paper.

The key result of Section III concerning the incoherent addition of failure probabilities
for different blocks, to lowest order, should be quite general:  it depends only on the 
base code being a nondegenerate distance 3 code and the interaction 
Hamiltonian being a sum of
pairwise products of elementary error operators for individual qubits.

The most important ``real-life'' complication which has been ignored here is, rather, 
the fact that, in practice, error
correction will not be a clean and purely mathematical operation, but a physical one, 
involving interacting systems and therefore itself subject to error.  The theory of
fault-tolerant error correction has shown that even when this is accounted for it 
should still be possible to achieve ``almost'' perfect error correction, provided that the 
error rate per gate is below a certain threshold, but it is here where one finds the 
(possibly) very small error thresholds which constitute the main challenge for large-scale
quantum computing.  

I am grateful to R. Laflamme and D. A. Meyer for comments.  This research has been sponsored 
by the National Science Foundation and by the Army Research Office.

\begin{figure}
\caption{The maximum failure probability for a qubit encoded with the plain 7-qubit code
(solid line) and with the same code concatenated once (dashed line) as a function of the 
product of interaction strength and interaction time}
\end{figure}

\begin{figure}
\caption{The maximum failure probability for a qubit encoded with the 5-qubit code
(solid line) and with the same code concatenated once (dashed line) as a function of the 
product of interaction strength and interaction time}
\end{figure}

\begin{figure}
\caption{Approximate results, valid for small $\delta t$, for the maximum failure probability
for a qubit encoded with the 7-qubit code concatenated 1 (no concatenation), 2, 3, and 4
levels deep (solid lines).  The exact results (from Figure 1) for the cases 1 and 2 are
shown as dashed lines.}
\end{figure}

\end{document}